\documentclass{aa}  

\usepackage{graphicx}
\usepackage[utf8]{inputenc}
\usepackage[T1]{fontenc}
\usepackage{caption}
\usepackage{subfigure}
\usepackage{graphics}
\usepackage{amsmath}
\usepackage{titlesec}
\usepackage{longtable}
\newcommand{\asec}{$^{\prime\prime}$}

\usepackage{txfonts}
\begin{document}

   \title{Seeds of Life in Space (SOLIS). XIII. Nitrogen fractionation towards the protocluster OMC-2 FIR4\thanks{NOEMA images are available in electronic form at the CDS via anonymous ftp to  cdsarc.u-strasbg.fr (130.79.128.5) or via http://cdsweb.u-strasbg.fr/cgi-bin/qcat?J/A+A/; SOLIS images will be provided by the Innovative Training Network (ITN) Astro-Chemical Origins (ACO) project as a deliverable.}}

   \author{L. Evans
          \inst{1,2}
          \and
          F. Fontani\inst{2}
          \and
          C. Vastel\inst{1}
          \and
          C. Ceccarelli\inst{3}
          \and
          P. Caselli\inst{4}
          \and
          A. L\'opez-Sepulcre\inst{3,5}
          \and
          R. Neri\inst{5}
          \and
          F. Alves\inst{4}
          \and
          L. Chahine\inst{5,6}
          \and
          C. Favre\inst{3}
          \and
          V. Lattanzi\inst{4}
          }

   \institute{IRAP, Université de Toulouse, 9 avenue du colonel Roche, 31028 Toulouse Cedex 4, France\\
              \email{levans@irap.omp.eu}
         \and
             INAF, Osservatorio Astrofisico di Arcetri, Largo E. Fermi 5, 50125 Firenze, Italy
         \and
            Univ. Grenoble Alpes, CNRS, Institut de Planetologie et d’Astrophysique de Grenoble (IPAG), 38000 Grenoble, France
        \and
            Max-Planck-Institut f\"{u}r extraterrestrische Physik (MPE), Giessenbachstrasse 1, 85748 Garching, Germany
        \and
            Institut de Radioastronomie Millimetrique (IRAM), 300 rue de la Piscine, Domaine Universitaire de Grenoble, 38406, Saint-Martin d’Hères, France
        \and
            Ecole doctorale de Physique, Universit\'{e} Grenoble Alpes, 110 Rue de la Chimie, 38400 Saint-Martin-d'H\`{e}res, France
             }


 
  \abstract
   {Isotopic fractionation is an important tool for investigating the chemical history of our Solar System. In particular, the isotopic fraction of nitrogen ($^{14}$N/$^{15}$N) is lower in comets and other pristine Solar System bodies with respect to the value measured for the protosolar nebula, suggesting a local chemical enrichment of $^{15}$N during the formation of the Solar System. Therefore, interferometric studies of nitrogen fractionation in Solar System precursors are needed for us to obtain clues about our astrochemical origins.}
   {In this work we have investigated the variation in the $^{14}$N/$^{15}$N ratio in one of the closest analogues of the environment in which the Solar System was born: the protocluster OMC-2 FIR4. We present the first comparison at high angular resolution between HCN and N$_2$H$^+$ using interferometric data.}
   {We analysed observations of the HCN isotopologues H$^{13}$CN and HC$^{15}$N in the OMC-2 FIR4 protocluster. Specifically, we observed the transitions H$^{13}$CN ($1-0$) and HC$^{15}$N ($1-0$) with the NOrthern Extended Millimeter Array (NOEMA) within the context of the IRAM Seeds Of Life In Space (SOLIS) Large Program. We combined our results with analysis of archival data  obtained with the Atacama Large Millimeter Array (ALMA) of N$_2$H$^+$ and its $^{15}$N isotopologues.}
   {Our results show a small regional variation in the $^{14}$N/$^{15}$N ratio for HCN, from $\sim$250 to 500. The ratios in the central regions of FIR4, where the candidate protostars are located, are largely consistent with one another and within that range ($\sim$300). They also show little variation from the part of the protocluster known to harbour a high cosmic-ray ionisation rate to the portion with a lower rate.  We also found 
    a small variation in the $^{14}$N/$^{15}$N ratio of N$_2$H$^+$ across different regions, from $\sim$200 to $\sim$400.}
   { These results suggest that local changes in the physical parameters occurring on the small linear scales probed by our observations in the protocluster do not seem to affect the $^{14}$N/$^{15}$N ratio in either HCN or N$_2$H$^+$ and hence that this is independent of the molecule used. 
   Moreover, the high level of irradiation due to cosmic rays does not affect the N fractionation either.}


   \maketitle
%

\section{Introduction}\label{Introduction}
The variety in the $^{14}$N/$^{15}$N ratios found across the Solar System (SS) and beyond represents one of the biggest mysteries in astrochemistry. The $^{14}$N/$^{15}$N ratio varies broadly among the objects of the SS itself, from approximately 140 in comets (\citet{Manfroid2009}; \citet{Mumma2011}; \citet{Shinnaka2014}), to 50-300 in meteorites (\citet{Bonal2010}; \citet{Aleon2010}), up to approximately 450 for the solar wind and Jupiter (\citet{Marty2011}). This last value is believed to represent the protosolar nebula (PSN) value (\citet{Furi2015}), indicating that comets and other pristine SS small bodies are enriched in $^{15}$N. However, the causes of this enrichment and, in particular, its relation to the chemical evolution of the PSN are not yet understood.

From the theoretical point of view, a former, popular explanation for the possible $^{15}$N enrichment was the low temperature isotope exchange reactions (\citet{Terzieva2000}, \citet{Rodgers2008}), similar to those at the origin of molecular deuterium enrichment in cold dense cores (e.g. \citet{Ceccarelli2014a} and references therein). However, recent chemical models have challenged this scenario due to the discovery of entrance barriers, which make these reactions unlikely to occur at low temperatures (\citet{Roueff2015}, \citet{Wirstrom2018}, \citet{Loison2019}). The new models also rule out significant nitrogen fractionation processes in the most abundant molecules during the chemical evolution of a star-forming core. In fact, significant variations in nitrogen fractionation are theoretically predicted only in extragalactic environments strongly affected by very high fluxes of cosmic rays (\citet{Viti2019}). 

Those predictions are at odds with the large variations in $^{14}$N/$^{15}$N measured in star-forming cores. In fact, the bulk $^{14}$N/$^{15}$N ratio in the present-day local interstellar medium has been calculated to be approximately 300 (\citet{Romano2017}, \citet{Adande2012}, \citet{Colzi2018a}), but large differences from the local interstellar value have been measured towards nearby pre- and proto-stellar objects, which seems to depend even on the molecule used (e.g. \citet{Womack1992}, \citet{Bizzocchi2013}, \citet{Hily-Blant2013, Hily-Blant2017, Hily-Blant2020}, \citet{Daniel2013, Daniel2016}, \citet{Guzman2017}, \citet{Redaelli2018a}, \citet{DeSimone2018}). The spread in values becomes even larger when measured towards high-mass star-forming cores (e.g. \citet{Fontani2015}, \citet{Zeng2017}, \citet{Colzi2018a, Colzi2018b}). These are located in different parts of the Galaxy and hence, in principle, are associated with elementary $^{14}$N/$^{15}$N ratios that are different from the local one due to the nitrogen isotopic galactocentric trend (\citet{Adande2012}, \citet{Colzi2018b}) expected from nucleosynthesis models (\citet{Romano2017, Romano2019}). However, the measured isotopic fractions cannot be explained only on the basis of the galactocentric trend (\citet{Colzi2018b}, \citet{Fontani2021}), indicating that local fractionation processes not yet included in the models may be at work. 

Moreover, most of the observations performed so far have been obtained with single-dish telescopes, thus providing average values of the $^{14}$N/$^{15}$N ratio over angular scales that can contain relevant variations in gas temperature and density. All this makes it challenging to identify whether the portions of the sources responsible for local enrichments (or depletions) in $^{15}$N correspond to particular ranges of physical conditions (in particular, density, temperature and extinction).

In fact, the few follow-up observations performed at high angular resolution have suggested local gradients in the $^{14}$N/$^{15}$N ratio. For example, \citet{Colzi2019} observed the high-mass protocluster IRAS 05358+3543 with a linear resolution of $\sim 0.05$~pc, or $\sim 10000$ au, finding that the $^{14}$N/$^{15}$N ratio in N$_2$H$^+$ shows an enhancement of a factor of $\sim 2$ (from $\sim 100 - 220$ to $\geq 200$) going from the inner dense core region to the diffuse, parsec-scale envelope; these results were interpreted as being a consequence of selective photodissociation (\citet{Heays2014}, \citet{Lee2021}). This mechanism, which is due to the self-shielding of $^{14}$N$_2$, predicts an increase in the $^{14}$N/$^{15}$N ratio in N$_2$H$^+$
in regions exposed to external UV irradiation, as indeed found by \citet{Colzi2019}). This finding has been confirmed in a sample of infrared-dark cloud cores by \citet{Fontani2021}, in which, again, the $^{14}$N/$^{15}$N ratio measured from N$_2$H$^+$ 
increases in the external envelope of the dense cores.

Local changes in the $^{14}$N/$^{15}$N ratio explainable by selective photodissociation have also been revealed in protoplanetary disks (\citet{Guzman2017}). In this case, the tracer used is HCN and a decrease in the $^{14}$N/$^{15}$N ratio has been found in the external layers directly illuminated by the radiation field of the central star. This different behaviour between N$_2$H$^+$ and HCN with respect to selective photodissociation is expected since the $^{15}$N-bearing species of N$_2$H$^+$ 
are formed from the molecular form of $^{15}$N, $^{14}$N$^{15}$N (dissociated at low extinctions), while those of HCN are formed from atomic $^{15}$N (enhanced at low extinctions). Deviations from these predictions have been proposed by \citet{Furuya2018}, who claim the $^{15}$N fractionation of N$_2$H$^+$ to be as low in the dense regions of pre-stellar cores as it is in the outer parts of the core, albeit for different reasons; in the dense regions it is because $^{15}$N atoms are transformed into $^{15}$NH$_3$ on the surface of dust grains and lost to the gas phase, so fewer $^{15}$N atoms will be available in the inner regions to form N$^{15}$NH$^+$ and $^{15}$NNH$^+$, while in outer parts this low fractionation is due to exposure to UV photons.

On the other hand, in very embedded, non-irradiated regions, the $^{14}$N/$^{15}$N ratio should not change significantly, something that has been recently confirmed by observations with a linear resolution of $\sim 600$ au (i.e. at core scales) obtained with the Atacama Large Millimeter Array (ALMA) of the protocluster OMC-2 FIR4, which have revealed negligible variations in the $^{14}$N/$^{15}$N ratio in the embedded protocluster cores (\citet{Fontani2020}). However, N$_2$H$^+$ is the only species investigated so far in $^{14}$N/$^{15}$N at core scales, therefore more observational results are needed to shed light onto the $^{15}$N fractionation process in cluster-forming environments.

In order to better investigate this, we obtained observations of isotopologues of HCN towards OMC-2 FIR4 in order to provide the first high angular resolution comparison of the $^{14}$N/$^{15}$N ratio between HCN and N$_2$H$^+$ using interferometric data in one of the best analogues of the environment in which our SS was born. The $^{14}$N/$^{15}$N ratio has already been investigated in our target at low angular resolution in multiple molecular species (\citet{Kahane2018}). This work will prepare the observational ground for direct comparisons with both objects of the SS (in particular with pristine small bodies as proxies of the early SS) and with chemical models.
In Sect. \ref{Presentation} we describe our target, the protocluster OMC-2 FIR4. The observations are presented in Sect. \ref{Observations}, the methods are described in Sect. \ref{Methods} and the results are introduced in Sect. \ref{Results} and discussed further in Sect. \ref{Discussion}. The main conclusions are given in Sect. \ref{Conclusions}.
\section{Source background}
\label{Presentation}
OMC-2 FIR4 is a protocluster located in the Orion Molecular Cloud 2 (OMC-2), at a distance of approximately 393$\pm$25 pc (\citet{Grossschedl2018}). The source consists of several continuum sources embedded in an extended, diffuse envelope (\citet{Shimajiri2008}, \citet{Lopez-Sepulcre2013}, \citet{Kainulainen2017}, \citet{Tobin2019}, Neri et al. in prep.) and it is part of a long, pc-scale filament containing several young stellar objects (YSOs) and star-forming regions (e.g. \citet{Furlan2016}, \citet{Hacar2018}). There are two neighbouring far-infrared (FIR) sources: one to the south-east (FIR5), which is not included in our observations; and one to the north-west (FIR3), which is included in our observations. FIR3 itself is a Class I YSO (\citet{Tobin2019}). The envelope of OMC-2 FIR4, having an angular size of $\sim 10-15$\asec or $\sim$ 4000-6000 au (e.g. \citet{Fontani2017, Fontani2020}), is believed to have a temperature of 35-45 K and an average H$_2$ volume density of 1.2$\times$10$^6$ cm$^{-3}$ (\citet{Ceccarelli2014b}), while FIR3, more compact ($\sim 5$\asec or $\sim$ 2000 au), is believed to have a higher temperature because it is likely associated with a hot corino (\citet{Tobin2019}, Ceccarelli et al. in prep.). OMC-2 FIR4 is believed to be one of the closest analogues to the early SS, due to its protocluster nature and evidence of external irradiation by an enhanced amount of energetic particles (\citet{Chaussidon2006}, \citet{Ceccarelli2014b}), probably accelerated by YSOs (\citet{Padovani2016}). Increasing evidence has been found to suggest that the Sun was born in a similar protocluster environment (\citet{Adams2010}) that was subject to irradiation of a very similar dose of energetic particles (\citet{Gounelle2013}, \citet{Ceccarelli2014b}). As such, it is an important object to study as direct comparisons can be made with what we know of the early stages of our own SS in order to learn more about our astrochemical origins.

Multiple previous works studying OMC-2 FIR4 (\citet{Ceccarelli2014b}, \citet{Fontani2017}, \citet{Favre2018}) have concluded with remarkable agreement that the cosmic-ray ionisation rate is enhanced in this source. In particular, \citet{Fontani2017} concluded from studying cyanopolyynes that a gradient in the cosmic-ray ionisation rate exists across the source, with the eastern portion of the protocluster having an enhanced cosmic-ray ionisation rate of $\zeta \sim 10^{-14}$ s$^{-1}$. Moreover, that portion of the source seems to be more exposed to energetic phenomena such as outflows from FIR3 and the interior of FIR4 (e.g. \citet{Osorio2017}, Lattanzi et al. in prep.).
\section{Observations}
\label{Observations}
\begin{figure}[ht]
\centering
\captionsetup{justification=centering}
\includegraphics[width=230pt]{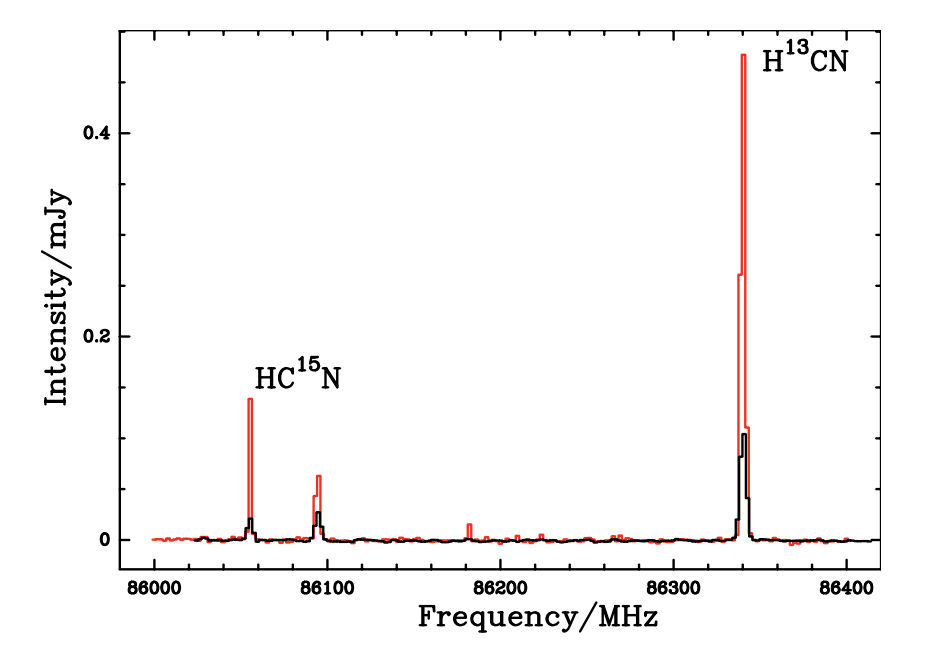}
\caption{Comparison between the flux recovered towards OMC-2 FIR4 by single-dish IRAM 30m observations (red) and towards the region equivalent to the 30m primary beam within the interferometric IRAM NOEMA observations of the same source (black). The figure shows that we recover approximately 20\% of the flux in both the H$^{13}$CN line and the HC$^{15}$N lines. The line at 86.1GHz is SO ($2_2-1_1$).}
\label{fig.flux}
\end{figure}

Our observations were obtained using the IRAM NOrthern Extended Millimeter Array (NOEMA) Interferometer as part of the Seeds Of Life In Space (SOLIS) Large Program (\citet{Ceccarelli2017}) with eight antennas on three days: April 29 and October 26, 2016 (both in C configuration) and January 16, 2017 (in A configuration). The phase centre was R.A.(J2000)=05$^h$35$^m$26$^s$97, Dec.(J2000)=--05$^{\circ}$09$^{\prime}$56$^{\prime \prime}$8 and the local standard of rest velocity was set to $11.4$ km s$^{-1}$. Baselines range from 20 to 760 m, providing an angular resolution of $\sim 4.1 \times 2.2$\asec (A+C configuration), corresponding to a linear scale of $\sim$ 860 au. The primary beam is $\sim 59.3$\asec at the representative frequency of 85~GHz. The system temperature, T$_{\rm sys}$, was typically in between $\sim 60$ and $\sim 100$~K in all tracks and the amount of precipitable water vapour was generally $\leq 5$~mm. The calibration of the bandpass and absolute flux scale was performed on 3C454.3 and MWC349 ($\sim 1.05$~Jy at 85.0~GHz), while 0524+034 and 0539--057 were used for calibration of the gains in phase and amplitude.
\\
\\
The lines were all observed with the Widex band correlator, providing a velocity resolution of 6.9 km s$^{-1}$. The continuum was imaged by averaging the line-free channels of the Widex and Narrow correlator units and subtracted from the spectrum in the ({\it u,v})-domain. Calibration and imaging were performed using the CLIC and MAPPING softwares of the GILDAS\footnote{The GILDAS software is developed at the IRAM and the Observatoire de Grenoble and is available at http://www.iram.fr/IRAMFR/GILDAS} package using standard procedures. The continuum image was self-calibrated and the solutions were applied to the lines.
\\
\\
In order to estimate the amount of flux that is filtered out by the interferometer, we compare our spectra of HCN isotopologues to the single-dish spectra of the same lines; we used publicly available IRAM 30m observations of OMC-2 FIR4 to achieve this (\citet{Kahane2018}). To this end, we extracted the HCN lines from the NOEMA cube within a polygon corresponding to the beam size of the single-dish observations (i.e.$~\sim$ 28\asec); this was compared to the IRAM 30m spectrum that had been resampled in velocity to match the resolution of our NOEMA observations (i.e.$~\sim 6.9$ km s$^{-1}$). The resulting comparison can be seen in Fig. \ref{fig.flux} and showed that we recovered approximately 20\% of the flux from both the H$^{13}$CN and HC$^{15}$N observations. As a result of the homogeneous effect of flux filtering on both lines, this caveat will have negligible impact on the calculated $^{14}$N/$^{15}$N ratios, assuming that their distribution is similar. Moreover, this missing flux likely affects mostly the emission from the envelope rather than that from the dense cores, where the emission of H$^{13}$CN and HC$^{15}$N is concentrated, as we show later in this section.

For the purpose of comparison, we also utilised archival data consisting of N$_2$H$^+$ ($1-0$) emission along with emission of both of its $^{15}$N isotopologues ($^{15}$NNH$^+$ ($1-0$) and N$^{15}$NH$^+$ ($1-0$)) towards OMC-2 FIR4 (ALMA project 2016.0.00681.S PI: F. Fontani). These lines were observed using ALMA in Band 3 (3~mm) with a velocity resolution of 0.2 km s$^{-1}$ and a spatial resolution of about 600 au. Technical details of these observations (e.g. calibration, configurations used, weather conditions, sensitivity achieved) are given in \citet{Fontani2020}. In the same way as for the HCN observations, the effect of flux filtering is the same for all isotopologues of N$_2$H$^+$, meaning that the effect on the $^{14}$N/$^{15}$N ratios is minimal.
\begin{table*}[!h]
\captionsetup{justification=centering}
\caption{Observational and spectral parameters of H$^{13}$CN and HC$^{15}$N as well as the HF components of N$_2$H$^+$ and its $^{15}$N isotopologues; all spectroscopic parameters have been taken from the CDMS (\citet{Endres2016})\\}
\label{hyperfine}      
\centering          
\begin{tabular}{c c c c c c}      
\hline
Transition & Rest Frequency & Quantum numbers$^{(a)}$ & Statistical Weight & E$_u$ & A$_{ul}$\\
 & GHz & & & K & s$^{-1}$\\
\hline \hline
\\
H$^{13}$CN & 86.3399214 & $1\rightarrow0$ & 9 & 4.14 & 2.23$\times$10$^{-5}$\\
\\
\hline
\\
HC$^{15}$N & 86.0549664 & $1\rightarrow0$ & 3 & 4.13 & 2.20$\times$10$^{-5}$\\
\\
\hline
\\
$^{15}$NNH$^+$ & 90.2634867 & $1~1$ $\rightarrow$ $0~1$ & 3 & 4.33 & 3.30$\times$10$^{-5}$\\ 
\\
 & 90.2639119 & $1~2$ $\rightarrow$ $0~1$ & 5 &  & \\
\\
 & 90.2645041 & $1~0$ $\rightarrow$ $0~1$ & 1 &  & \\
\\
\hline
\\
N$^{15}$NH$^+$ & 91.2042615 & $1~1$ $\rightarrow$ $0~1$ & 3 & 4.38 & 3.44$\times$10$^{-5}$\\
\\
 & 91.2059911 & $1~2$ $\rightarrow$ $0~1$ & 5 &  & \\
\\
 & 91.2085167 & $1~0$ $\rightarrow$ $0~1$ & 1 &  & \\
\\
\hline
\\
N$_2$H$^+$ & 93.1716157 & $1~1~0$ $\rightarrow$ $0~1~1$ & 1 & 4.47 & 3.63$\times$10$^{-5}$\\
\\
 & 93.1719106 & $1~1~2$ $\rightarrow$ $0~1~2$ & 10 &  & 3.60$\times$10$^{-5}$\\
\\
 & 93.1720477 & $1~1~1$ $\rightarrow$ $0~1~0$ & 9 &  & 3.63$\times$10$^{-5}$\\
\\
 & 93.1734734 & $1~2~2$ $\rightarrow$ $0~1~1$ & 10 &  & 3.63$\times$10$^{-5}$\\
\\
 & 93.1737699 & $1~2~3$ $\rightarrow$ $0~1~2$ & 7 &  & 3.63$\times$10$^{-5}$\\
\\
 & 93.1739640 & $1~2~1$ $\rightarrow$ $0~1~1$ & 9 &  & 3.63$\times$10$^{-5}$\\
\\
 & 93.1762595 & $1~0~1$ $\rightarrow$ $0~1~2$ & 9 &  & 3.63$\times$10$^{-5}$\\
\\
\hline
\end{tabular}
\\
\footnotesize{$^{(a)}$The quantum numbers represent the J (H$^{13}$CN and HC$^{15}$CN), J and $F_1$ ($^{15}$NNH$^+$ and N$^{15}$NH$^+$) and J, $F_1$ and F (N$_2$H$^+$) levels}
\end{table*}

\begin{figure}[ht]
\centering
\captionsetup{justification=centering}
\begin{subfigure}{}
\includegraphics[width=230pt]{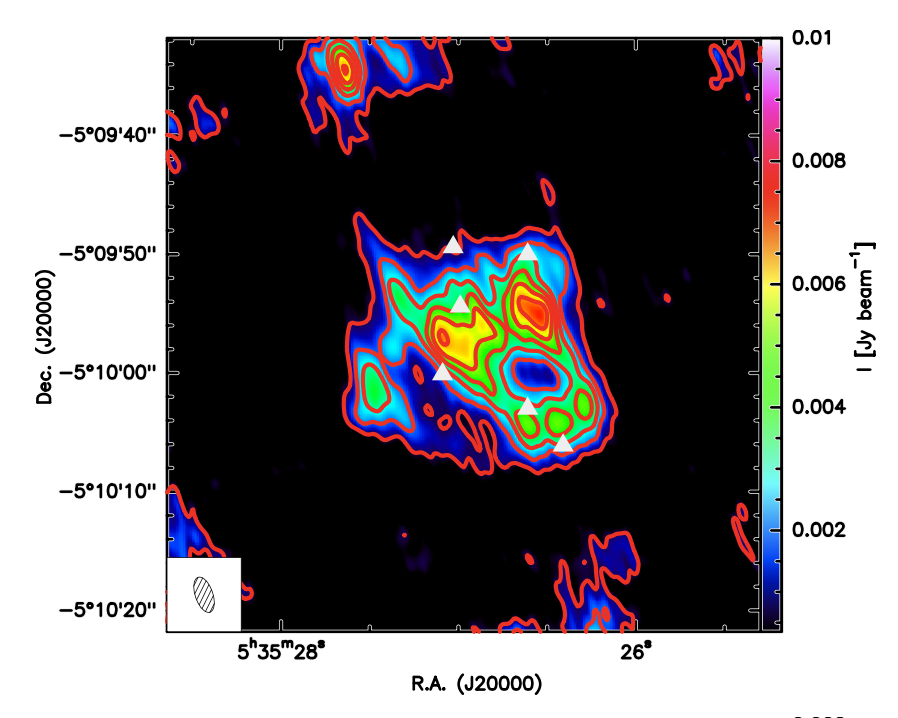}
\end{subfigure}
\begin{subfigure}{}
\includegraphics[width=230pt]{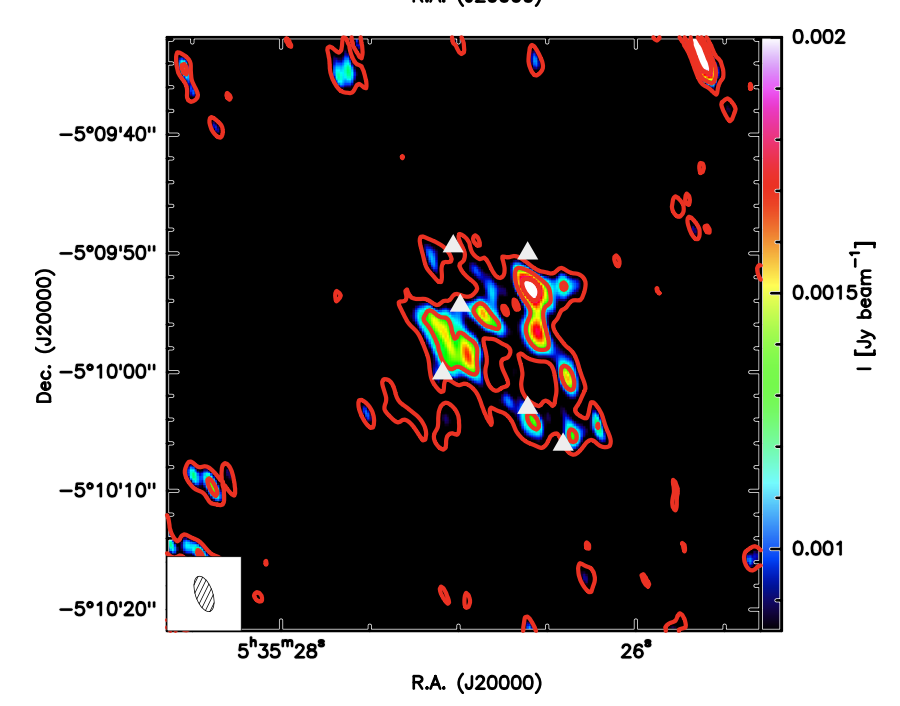}
\end{subfigure}
\caption{Upper panel: H$^{13}$CN ($1-0$) average emission map in colour scale and red contours. Contours start at the level of 4$\sigma$, corresponding to 0.70 mJy beam$^{-1}$ (1$\sigma$ = 0.18 mJy beam$^{-1}$) and are in intervals of 6$\sigma$, corresponding to 1.17 mJy beam$^{-1}$. Lower panel: HC$^{15}$N ($1-0$) average emission map in colour scale and red contours. Contours start at the level of 4$\sigma$, corresponding to 0.84 mJy beam$^{-1}$ (1$\sigma$ = 0.21 mJy beam$^{-1}$) and are in levels of 4$\sigma$, corresponding to 0.84 mJy beam$^{-1}$. Continuum sources identified by Neri et al. (in prep.) are marked by white triangles. The wedge on the right indicates the range of flux density (Jy beam$^{-1}$). The ellipse in the bottom-left corner of each panel represents the NOEMA synthesised beam.}
\label{contours}
\end{figure}

\begin{figure}[ht]
\centering
\captionsetup{justification=centering}
\begin{subfigure}{}
\includegraphics[width=230pt]{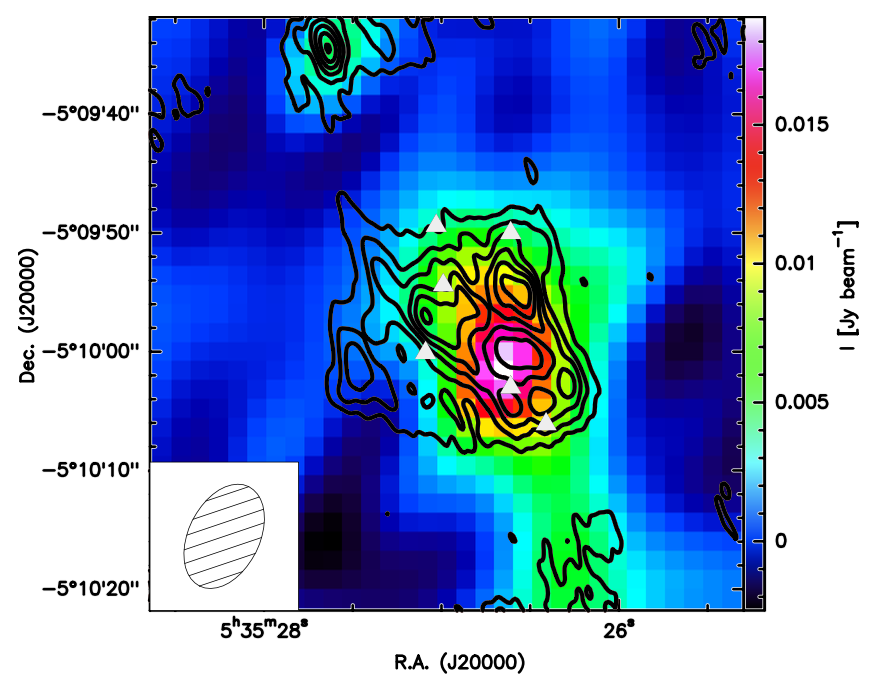}
\end{subfigure}
\begin{subfigure}{}
\includegraphics[width=230pt]{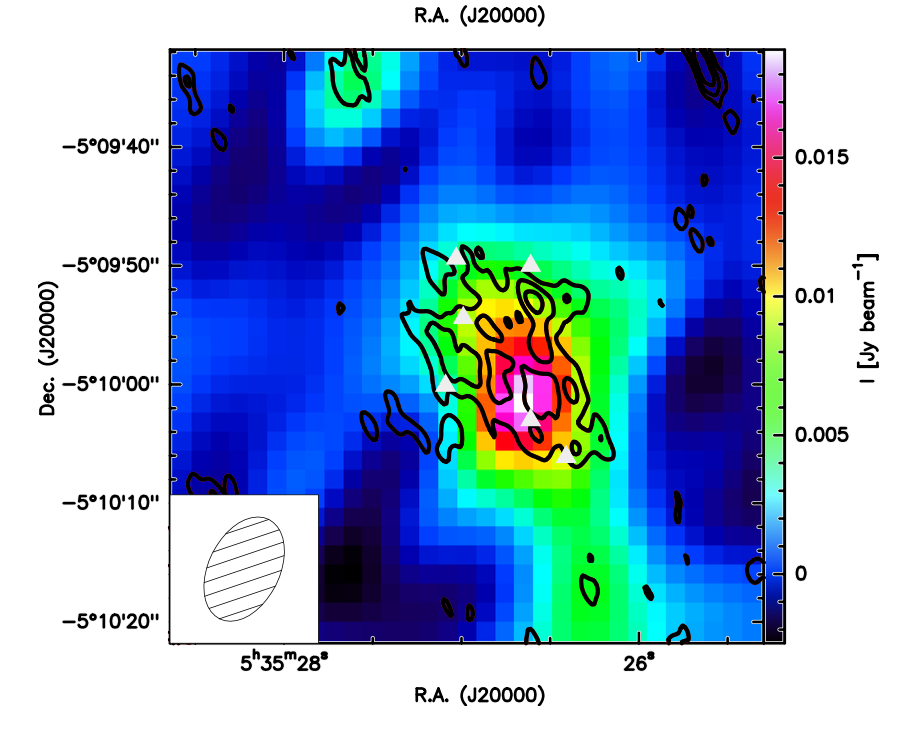}
\end{subfigure}
\caption{Upper panel: Continuum emission map at 3~mm (from \citet{Fontani2017}) in colour scale with H$^{13}$CN ($1-0$) emission overlaid in black contours. Contours start at the level of 4$\sigma$, corresponding to 0.70 mJy beam$^{-1}$ (1$\sigma$ = 0.18 mJy beam$^{-1}$) and are in intervals of 6$\sigma$, corresponding to 1.17 mJy beam$^{-1}$. Lower panel: Same as the upper panel but the black contours show the HC$^{15}$N ($1-0$) average emission. Contours start at the level of 4$\sigma$, corresponding to 0.84 mJy beam$^{-1}$ (1$\sigma$ = 0.21 mJy beam$^{-1}$) and are in levels of 4$\sigma$, corresponding to 0.84 mJy beam$^{-1}$. Continuum sources identified by Neri et al. (in prep.) are marked by white triangles. The wedge on the right indicates the range of flux density (Jy beam$^{-1}$). The ellipse in the bottom-left corner of each panel represents the NOEMA synthesised beam of the 3~mm continuum emission.}
\label{continuum}
\end{figure}

Table \ref{hyperfine} shows the spectral parameters of the observed lines, including the parameters for the hyperfine (HF) components of N$_2$H$^+$ and its isotopologues. 
Figure \ref{contours} shows maps of the emission of both the H$^{13}$CN and HC$^{15}$N ($1-0$) lines averaged over the channels with signal-to-noise ratio (S/N) larger than 3, corresponding to an approximate velocity range of -1 to 18km s$^{-1}$ for H$^{13}$CN and 5 to 15 km s$^{-1}$ for HC$^{15}$N. For comparison, Fig. \ref{continuum} shows the continuum emission obtained from the NOEMA data at 3~mm published in \citet{Fontani2017}. It can be seen from these maps that the H$^{13}$CN ($1-0$) emission is ubiquitously stronger than the HC$^{15}$N ($1-0$) emission across all regions of both FIR4 and FIR3. H$^{13}$CN and HC$^{15}$N show similar peak locations, with both having a peak in FIR3 and the north-western and north-eastern regions of the central part of FIR4. Three peaks in the southern inner region appear in HC$^{15}$N and are faintly seen in H$^{13}$CN among generally stronger surrounding emission. The most notable difference between the two maps is the presence of an eastern peak in the outer region of FIR4 in the H$^{13}$CN map while this is absent in the HC$^{15}$N map. It should also be noted that there is a hole in the H$^{13}$CN (and possibly HC$^{15}$N) emission towards the central region of FIR4 that is absent from the continuum emission and from any molecular line studied so far (e.g. \citet{Fontani2017, Fontani2020}, \citet{Favre2018}).

\begin{figure}[ht]
\centering
\captionsetup{justification=centering}
\begin{subfigure}{}
\includegraphics[width=230pt]{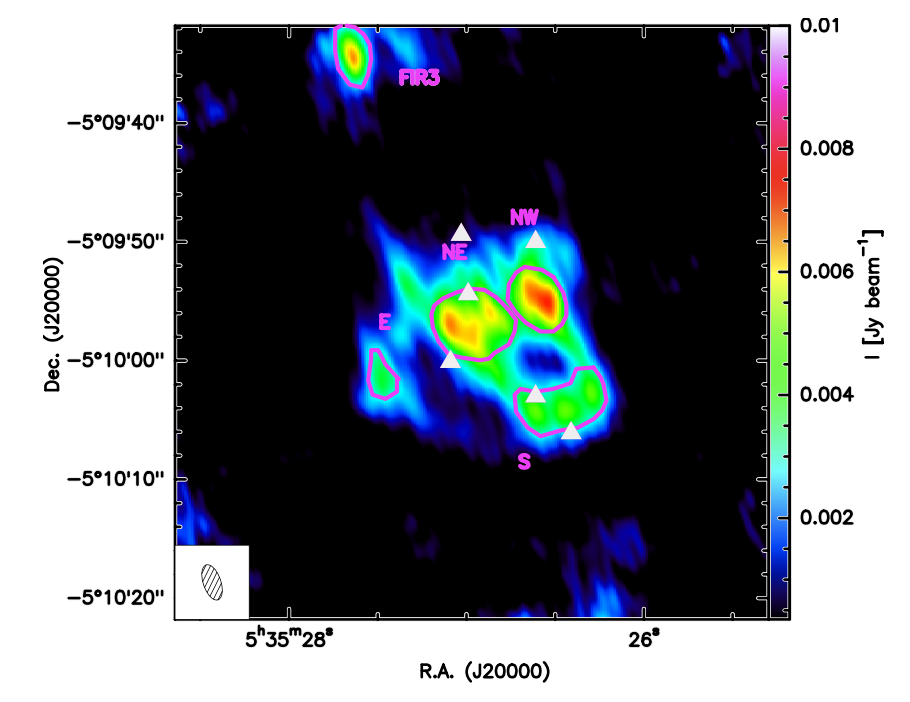}
\end{subfigure}
\begin{subfigure}{}
\includegraphics[width=230pt]{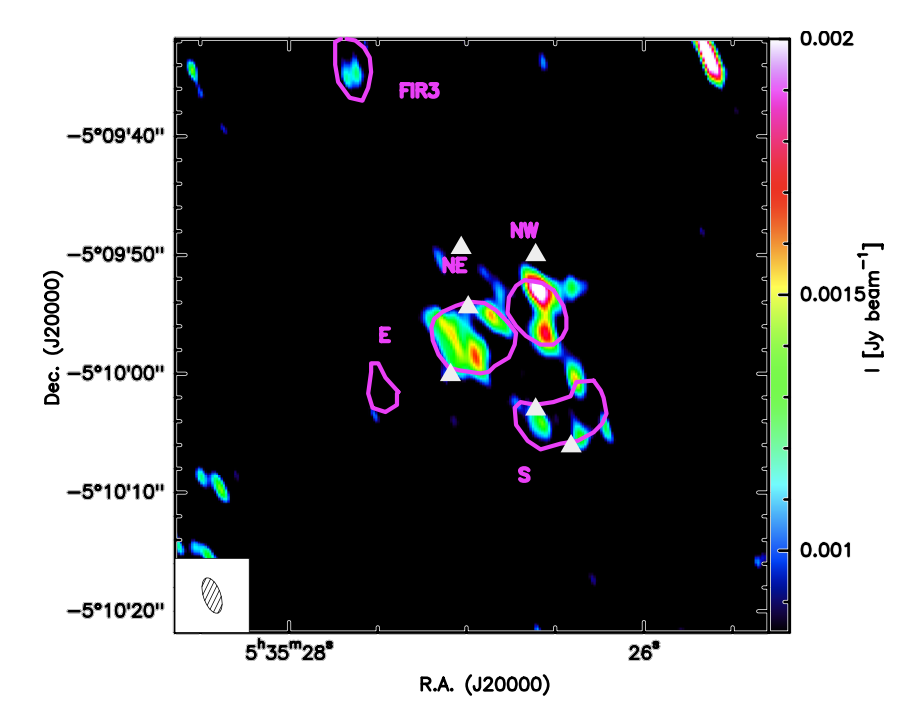}
\end{subfigure}
\caption{Upper panel: H$^{13}$CN ($1-0$) average emission map with extraction regions enclosed by pink curves. Lower panel: HC$^{15}$N ($1-0$) average emission map with extraction regions enclosed by pink curves. The wedge on the right of each panel indicates the range of flux density (Jy beam$^{-1}$). The ellipse in the bottom-left corner of each panel represents the NOEMA synthesised beam. White triangles mark the positions of continuum sources as identified by Neri et al. (in prep.).}
\label{fig_1}
\end{figure}

The H$^{13}$CN ($1-0$) emission map suggests the presence of five gaseous cores in which the average emission is above 3$\sigma$ rms (see Fig. \ref{fig_1}). One is peaked on FIR3, the rest throughout FIR4. These sub-structures were chosen as the regions from which the spectra will be extracted and the isotopic $^{14}$N/$^{15}$N ratio for HCN will be calculated.

As stated above, we extracted spectra of N$_2$H$^+$, $^{15}$NNH$^+$ and N$^{15}$NH$^+$ $(1-0)$, observed with ALMA, from the dataset published in \citet{Fontani2020}. The analysis of these spectra will allow us to compare, for the first time at high angular resolution, the $^{14}$N/$^{15}$N ratio simultaneously in HCN and N$_2$H$^+$ (i.e. the two molecules in which nitrogen fractionation should behave in an opposite way) if both are regulated by selective photodissociation. Figure \ref{fig.spectra} in the Appendix shows the spectra that resulted from our extraction from the chosen regions. In the case of FIR3, the spectrum shows no detection of N$_2$H$^+$ emission. This could be due to the destruction of this molecule by CO or it could be that the emission is so faint and extended that it has been resolved out by the interferometer. In the case of the $^{15}$N isotopologues, $^{15}$NNH$^+$ also shows no emission towards FIR3, but there does appear to be a detection of N$^{15}$NH$^+$ as a clear peak is present in this spectrum. This suggests that the reason that the main N$_2$H$^+$ isotopologue is undetected is because the emission is resolved out by our interferometer as it is too extended; the presence of N$^{15}$NH$^+$ is explainable by the fact that this emission is likely to be more compact than the main isotopologue and so is not resolved out. For this reason, the $^{14}$N/$^{15}$N ratio in FIR3 from N$_2$H$^+$ cannot be derived and will not be discussed further.

The N atom has a non-zero nuclear spin, which gives rise to hyperfine structure (HFS) splitting in spectra of N-bearing species, including HCN and N$_2$H$^+$, as well as the $^{15}$N isotopologues of the latter. In the case of H$^{13}$CN and HC$^{15}$N, the HFS is not resolved in our spectra due to the velocity resolution of these observations being insufficient to see the separate HFS components. However, in the spectra of N$_2$H$^+$ and its $^{15}$N isotopologues, the velocity resolution is sufficient to resolve the HFS (see Fig. \ref{fig.spectra}). In the case of the $^{15}$N isotopologues, we can resolve all three HF components, while in the case of the main isotopologue, we can observe seven HF components.
\section{Methods}\label{Methods}
\subsection{Extraction of the spectra and analysis}\label{Extraction}
Using the MAPPING package of GILDAS (see Sect. \ref{Observations}), the spectra were extracted in flux density units (Jy) from the five regions illustrated in Fig. \ref{fig_1}, which have been labelled depending on the position of the peak with respect to the phase centre, as: north-west (NW), north-east (NE), south (S), east (E) and FIR3. Due to the complexity of the emission and the partial overlap of some sub-structures, we decomposed the average emission map of H$^{13}$CN by eye into the polygons shown in Fig. \ref{fig_1}, which do not overlap and pick up the intensity peak of each sub-region. The CLASS package of GILDAS was employed to convert our spectra into temperature units (K) before they were imported into CASSIS\footnote{Based on analysis carried out with the CASSIS software (\citet{Vastel2015}) and JPL (http://spec.jpl.nasa.gov/) and CDMS (https://cdms.astro.uni-koeln.de/classic/) molecular databases. CASSIS has been developed by IRAP-UPS/CNRS (http://cassis.irap.omp.eu).}.

As can be seen in Fig. \ref{fig.spectra}, the lines of H$^{13}$CN and HC$^{15}$N show generally a Gaussian shape; hence, each line was fitted with a single Gaussian in CASSIS using Levenberg-Marquardt (L-M) fitting. The fit provides several line parameters, among which is the full width at half maximum (FWHM), which was then used as one of the input parameters for local thermal equilibrium (LTE) analysis using CASSIS. In order to check our LTE assumption, which is satisfied if the H$_2$ density is larger than the critical density of the analysed transition, we computed the critical density of each of our lines from the equation:
\begin{gather}
n_{\rm CRIT}=\frac{A_{\rm ul}}{C_{\rm ul}} .
\label{eq.criticaldensity}
\end{gather}
In Eq \ref{eq.criticaldensity}, $n_{\rm CRIT}$ represents the critical density (cm$^{-3}$) while $A_{\rm ul}$ (s$^{-1}$) represents the Einstein coefficient and $C_{\rm ul}$ (cm$^{-3}$s$^{-1}$) represents the collisional coefficient. We use collisional coefficients taken from the CASSIS collision database and calculated by \citet{Hernandez-Vera2014} for HCN and by \citet{Daniel2005} and \citet{Lique2015} for N$_2$H$^+$. For the $1-0$ transition we can assume that the $^{15}$N isotopologues (and $^{13}$C-isotopologues) have the same collisional coefficients as their respective main isotopologues. The critical density at 40 K was calculated to be 1.62$\times$10$^6$ cm$^{-3}$ for H$^{13}$CN, 1.59$\times$10$^6$ cm$^{-3}$ for HC$^{15}$N,  1.61$\times$10$^5$ cm$^{-3}$ for $^{15}$NNH$^+$, 1.66$\times$10$^5$ cm$^{-3}$ for N$^{15}$NH$^+$ and 1.77$\times$10$^5$ cm$^{-3}$ for N$_2$H$^+$. In the case of N$_2$H$^+$ and its isotopologues, we can safely assume LTE as these values for critical density are smaller than the average H$_2$ volume density of OMC-2 FIR4 (1.2$\times$10$^6$ cm$^{-3}$; \citet{Ceccarelli2014b}); for H$^{13}$CN and HC$^{15}$N, the calculated critical density values are the same order of magnitude as the volume density and so the approximation of LTE conditions is reasonable. It should be noted that we used the density value from \citet{Ceccarelli2014b} because it is an average value calculated over the whole system, making it the best comparison that we can make with available data.

Radiative transfer through gas with a constant excitation temperature (\textit{T$_{ex}$}) leads to the following expression for the resulting intensity as synthesised beam temperature $T_{\rm b}$($\nu$):
\begin{gather}
T_b(\nu)=\frac{h\nu}{k}\Bigg(\frac{1}{e^{\frac{h\nu}{kT_{\rm ex}}}-1}-\frac{1}{e^{\frac{h\nu}{kT_{\rm bg}}}-1}\Bigg)(1-e^{-\tau(\nu)}),
\label{eq.tau}
\end{gather}
where \textit{c}, \textit{h} and \textit{k} represent the speed of light, Boltzmann constant and Planck constant, respectively, in cgs units, while $\tau(\nu)$ is the opacity of the line. We take the background temperature (\textit{T$_{\rm bg}$}) as that of the cosmic microwave background (CMB; i.e. 2.73 K). The column density $N_{\rm TOT}$ (cm$^{-2}$) is then obtained directly from the measure of the opacity in the line centre computed in Eq. \ref{eq.tau} ($\tau_0$), for which \textit{T$_{\rm b}$} is a maximum:
\begin{gather}
N_{\rm TOT}=\frac{8\pi\nu^3(\Delta \rm{v})\tau_0 Q(T_{\rm ex})\sqrt{\pi}}{2\sqrt{ln2}}\frac{e^{\frac{E_{\rm u}}{kT_{\rm ex}}}}{c^3A_{\rm ul}g_{\rm u}(e^{\frac{h\nu}{kT_{\rm ex}}}-1)},
\label{eq.lte}
\end{gather}
where $\nu$ is frequency (Hz), {$\Delta\rm{v}$} represents the line width (cm s$^{-1}$), $Q(T_{\rm ex})$ is the partition function at excitation temperature $T_{\rm ex}$, $E_{\rm u}$ is the energy of the upper level (K), $A_{\rm ul}$ is the Einstein coefficient (s$^{-1}$) and $g_{\rm u}$ is the upper state degeneracy. Eqs.  \ref{eq.lte} (and \ref{eq.tau}), which can be found in the Formalism for the CASSIS software\footnote{http://cassis.irap.omp.eu/docs/RadiativeTransfer.pdf} (author: C. Vastel), describe how to compute the total column density (and opacity) from our spectroscopic and observational parameters under the conditions of LTE and optically thin lines. We justify our use of the optically thin assumption in Sects. \ref{HCN} and \ref{N2H+} for HCN and N$_2$H$^+$, respectively.
\subsection{HCN isotopologues}
\label{HCN}
Our analysis obtained opacities ranging from 8.1$\times$10$^{-4}$ to 1.6$\times$10$^{-2}$, enabling us to assume that the lines are optically thin and use the method described in Sect. \ref{Extraction}. Due to the fact that we only observed one transition for both H$^{13}$CN and HC$^{15}$N, the excitation temperature ($T_{\rm ex}$) cannot be derived directly from the data and it has to be assumed. Therefore, the spectra were modelled for two fixed temperatures, corresponding to the range of temperatures found for FIR4 using previous data: an upper temperature of 45 K and a lower temperature of 35 K (taken from \citet{Ceccarelli2014b}, \citet{Favre2018}). For FIR3, the upper temperature was fixed at 150 K and the lower temperature was fixed at 100 K (assuming that FIR3 is a hot corino; \citet{Tobin2019}, Ceccarelli et al. in prep.). The column density was repeatedly varied until the model matched the observations, taking into account a 10\% calibration error in the latter. Using this method, we obtained best-fit values (along with upper and lower limits) for the column densities of H$^{13}$CN and HC$^{15}$N in each region. It should be noted that H$^{13}$CN is potentially better than HCN for computing the $^{14}$N/$^{15}$N ratio because HCN is expected to be optically thick. However, the column density ratio $N_{\rm TOT}$(H$^{13}$CN)/$N_{\rm TOT}$(HC$^{15}$N) has to be corrected by the $^{12}$C/$^{13}$C ratio, which was assumed to be 50$\pm$5 as measured by \citet{Kahane2018} using CN, HCN and HC$_3$N lines. The $^{12}$C/$^{13}$C ratio measured from HCN specifically in this paper is tentative; the authors calculated their HCN-specific value of 36$\pm$15 by comparing HC$^{15}$N to H$^{13}$C$^{15}$N. They point out in this paper that due to the weakness (2.5$\sigma$) of the H$^{13}$C$^{15}$N detection, it is possible that the obtained $^{12}$C/$^{13}$C ratio is underestimated (see \citet{Kahane2018} for further discussion). Taking this into account, we decided to use the average value of 50$\pm$5. We also note that the uncertainty associated with the average $^{12}$C/$^{13}$C ratio calculated in \citet{Kahane2018} is included in our calculated uncertainties associated with $^{14}$N/$^{15}$N ratios of HCN.
\\
\\
\subsection{N$_2$H$^+$}
\label{N2H+}
\subsubsection{CLASS}\label{CLASS}
Due to the aforementioned presence of HFS, we first used the specific HFS fitting routine that is provided within CLASS to model our spectra. 
This routine requires the following inputs for each isotopologue: relative intensity, r$_i$ (with sum normalised to 1) and velocity separation from the reference component (computed from the Doppler law based on the frequency differences obtained in the lab), \textit{\rm{v}$_i$} (km s$^{-1}$) for each HF component. Both inputs were in this case obtained from the Cologne Molecular Database for Spectroscopy (CDMS) for each isotopologue. 
If we define the central velocity of the \textit{i}th component as \textit{\rm{v}$_{0,i}$=\rm{v}$_i$+p$_2$}, then 
we can calculate the opacity of our lines, $\tau$, using Eq. \ref{eq.opacity}:
\begin{gather}
\label{eq.opacity}
\tau(\nu)=p_4\Sigma^N_{i=1}r_ie^{-4ln2(\frac{\rm{v}-\rm{v}_i-p_2}{p_3})^2}.
\end{gather}
In this equation, \textit{$\tau$} represents the opacity as a function of frequency (\textit{$\nu$}), \textit{N} represents the number of HF components, \textit{r$_i$} is the relative intensity of the \textit{i}th HF component and \textit{\rm{v}$_i$} is the velocity separation of the \textit{i}th HF component from the reference component, (km s$^{-1}$). 
The fitting procedure described above provides optically thin lines, (see Table \ref{Table_error_hcn} for the opacity values obtained by CLASS HFS) and hence we can estimate $N_{\rm TOT}$ from the relation:
\\
\\
\begin{gather}
\label{eq.caselli}
N_{\rm TOT}=\frac{8\pi W \nu^3}{A_{\rm ul}c^3g_{\rm u}}\frac{Q(T_{\rm ex})e^{\frac{E_{\rm u}}{kT_{\rm ex}}}}{e^{\frac{h\nu}{kT_{\rm ex}}}-1}\frac{1}{J_{\nu}(T_{\rm ex})-J_{\nu}(T_{\rm bg})}.
\end{gather}
Equation \ref{eq.caselli} provides $N_{\rm TOT}$ from the total integrated intensity of the line ($W$) in the case of optically thin transitions and all rotational levels being characterised by a single $T_{\rm ex}$. The equivalent Rayleigh-Jeans temperature, 
$J_{\nu}(T)$, is defined as
\begin{gather}
\label{eq.Jv}
J_\nu(T_{ex,bg})=\frac{h\nu}{k}\frac{1}{e^{\frac{h\nu}{kT_{ex,bg}}}-1}.
\end{gather}
Since we have HFS and optically thin lines, we can calculate the total integrated intensity in two ways: when all HF components are detected, we sum the integrated intensity of all of them; when only the strongest HF component is detected, we multiply its integrated intensity by the known relative ratio with respect to the total.

For the $^{15}$N isotopologues, the opacity derived from the fit 
confirms that we have optically thin lines ($\tau \leq 0.1$) in all but one case: N$^{15}$NH$^+$ towards the NW region. In this line, the opacity of the sum of the components is $\sim 1.4$, but the error is so large (1.4 $\pm$ 1.4) that we can assume optically thin conditions also in this case, following \citet{Caselli2002}.
For the lines for which only the strongest HF component is detected, we know that this component has a relative intensity of $\frac{5}{9}$ compared to the total; therefore, we divided its integrated intensity by $\frac{5}{9}$ in order to obtain the total integrated intensity, $W$, for the $^{15}$N isotopologues.

On the other hand, the output 
of the CLASS HFS fitting routine tells us that the opacity of the sum of the components for the main isotopologue (N$_2$H$^+$) is larger than 1, meaning that in all cases we have optically thick lines. 
However, the faintest component, $J'F'-JF=01\rightarrow12$, has, according to the relative intensity ratio, an opacity that is $\frac{3}{27}$ of the total, providing opacities lower than 1 in all spectra. In addition, this component is well separated from the others in the spectrum, therefore its integrated intensity is not contaminated by emission from any other component.
We measured its integrated intensity and divided it by the factor $\frac{3}{27}$ to derive the total one.

Table \ref{Table_error_hcn} shows the best-fit FWHM and total integrated intensity values of our lines with uncertainties given by:
\begin{gather}
\Delta N_{\rm TOT}=N_{\rm TOT}\frac{\Delta W}{W+0.1}
\label{eq.error}
\end{gather}
\begin{gather}
\Delta W = \sqrt{(cal/100\times W)^2+(rms\sqrt{2\times FWHM\times \Delta\rm{v}_{res}})^2}.
\label{eq.areaerror}
\end{gather}
In these equations, $N_{\rm TOT}$ and $\Delta N_{\rm TOT}$ represent column density (cm$^{-2}$) and associated error, respectively, $W$ and $\Delta W$ represent the integrated area (K cm$^{-1}$) and associated error, respectively, \textit{cal} represents the calibration error (assumed in this case to be 10\%), $\Delta \rm{v_{res}}$ represents the velocity resolution (km s$^{-1}$) (in our case 0.2 km s$^{-1}$) and \textit{rms} is 10.1 mK for $^{15}$NNH$^+$, 10.3 mK for N$^{15}$NH$^+$ and 31 mK for N$_2$H$^+$. From fitting our spectra using L-M fitting in CASSIS, FWHM is obtained as 0.8 km s$^{-1}$. It should be noted that the values for $W$ obtained using the method described previously are given in K km s$^{-1}$; therefore, for Eq. \ref{eq.caselli}, these are converted to K cm s$^{-1}$ as everything else is in cgs units.

All spectroscopic parameters used in Eq. \ref{eq.caselli} are taken from the CDMS database. The CMB temperature (\textit{$T_{bg}$}) is assumed to be 2.73 K. For the excitation temperature ($T_{\rm ex}$), we cannot constrain this using our analysis alone because, despite resolving the HFS in our spectra, the opacity of the N$_2$H$^+$ spectra is not well constrained and the $^{15}$N-isotopologue lines are optically thin (see Table \ref{Table_error_hcn}). Therefore, we must take our values from previous measurements. We fixed $T_{\rm ex}$ to 35 K and 45 K (\citet{Ceccarelli2014b}, \citet{Favre2018}.
\subsubsection{CASSIS}
\label{CASSIS}
Within CASSIS, there is a Python library (LineAnalysisScripting.py) that allows users to create Jython scripts to model sources in LTE 
using the Markov chain Monte Carlo  (MCMC) method. As part of the script, we define a number of parameters that will be varied and give upper and lower limits for the variation range. In most cases, we have sufficient S/Ns to vary four parameters: \textit{N$_{TOT}$}, \textit{T$_{ex}$}, FWHM and \textit{\rm{V}$_{LSR}$}. From L-M fitting, we know the order of magnitude for \textit{N$_{TOT}$} in each case as well as the FWHM for all of the components ($\sim$0.8 km s$^{-1}$). Therefore, \textit{N$_{TOT}$} is varied between 10$^{10}$ cm$^{-2}$ and 10$^{15}$ cm$^{-2}$ for the $^{15}$N isotopologues and between 10$^{12}$ cm$^{-2}$ and 10$^{15}$ cm$^{-2}$ for the main isotopologue. From this we also vary FWHM between 0.3 km s$^{-1}$ and 1.2 km s$^{-1}$ in each case. For \textit{T$_{ex}$}, once again we use previous research of this source to determine our upper and lower limits: 50 K and 30 K, respectively (5 K above and below the upper and lower limits used to model the spectra and calculate \textit{N$_{TOT}$} from the CLASS method). Also from previous research, the accepted value for the \textit{\rm{V}$_{LSR}$} of OMC-2 FIR4 is 11.4 km s$^{-1}$ (e.g. \citet{Kahane2018}, \citet{Ceccarelli2017}); therefore, this parameter is varied between 8 and 15 km s$^{-1}$.

In the case of the $^{15}$N isotopologues, since we have lower S/N in general, we must constrain at least one of the parameters (in cases where we have higher S/N, we obtain good fits by leaving the parameters unfixed and allowing the software to find the best fit; if we have low S/N we cannot do this). Therefore, in these cases, we constrain the range for \textit{T$_{ex}$} to just 3$\sigma$ above and below the \textit{T$_{ex}$} value obtained for the main isotopologue in the corresponding position. Meanwhile, three of the spectra show very low S/N: the E Region of both $^{15}$N isotopologues as well as the S Region of N$^{15}$NH$^+$. For these cases, we also constrain the \textit{\rm{V}$_{LSR}$}, fixing this parameter to the value obtained in the corresponding position in the main isotopologue. In all cases, the script is run and the model is compared to the observations.

The values obtained with the two methods are the same within error bars except for the cases with higher opacity (i.e. NW Region) or with low S/N (i.e. E Region in both $^{15}$N isotopologues and S Region in N$^{15}$NH$^+$). For the latter, the fit obtained using MCMC with the aforementioned limits was not optimal and underestimated \textit{N$_{TOT}$} with respect to the observations, causing overestimations of the $^{14}$N/$^{15}$N ratio. Therefore, we concluded that these three spectra had insufficient S/N to be analysed using MCMC; therefore, to analyse these data, first we fitted a single Gaussian to the strongest HF component using L-M fitting in the relevant spectrum (see Table \ref{Table_error_hcn}) to obtain a value of \textit{\rm{V}$_{LSR}$} for each spectrum. We used this value to fix \textit{\rm{V}$_{LSR}$} in a line analysis while fixing \textit{T$_{ex}$} to 35 K and 45 K to give upper and lower limits. The \textit{N$_{TOT}$} was then varied manually until good fits were obtained, taking into account a 15\% calibration error. The best-fit values in all cases for \textit{T$_{ex}$}, FWHM and \textit{\rm{V}$_{LSR}$} are in Table \ref{Table_error_hcn}. Table \ref{Table_ncol_hcn} shows the column density values for H$^{13}$CN and HC$^{15}$N that were obtained from the method described above along with the resultant $^{14}$N/$^{15}$N ratios. The resultant values of \textit{N$_{TOT}$} for N$_2$H$^+$ and its isotopologues (Table \ref{Table_ncol_n2h+}) were used to calculate the $^{14}$N/$^{15}$N ratio for N$_2$H$^+$ (Table \ref{Table_2}). 
\begin{table*}[!h]
\captionsetup{justification=centering}
\caption{Values for the parameters used in Eqs.  \ref{eq.error} and \ref{eq.areaerror}, plus those varied during MCMC modelling for N$_2$H$^+$ and its $^{15}$N isotopologues. Calibration errors of 10\%  and 15\% were used in calculations involving the CLASS and MCMC models, respectively. We note that, due to the low spectral resolution, the HFS structure cannot be constrained for HCN isotopologues and so the spectra are fit with one transition only, the FWHM of which is included in this table for each region. Also, for this reason, \rm{V}$_{LSR}$ is fixed to the accepted source value of 11.4 km s$^{-1}$. The RMS was 5.1 mK for H$^{13}$CN, 4.4 mK for HC$^{15}$N, 10.1 mK for $^{15}$NNH$^+$, 10.3 mK for N$^{15}$NH$^+$ and 31 mK for N$_2$H$^+$.}
\label{Table_error_hcn} 
\centering
\begin{tabular}{c c c c c c c}
\hline  
Line & Region & FWHM & \rm{V}$_{LSR}$ & T$_{ex}$ & Total Integrated Area & Opacity\\
 & & km s$^{-1}$ & km s$^{-1}$ & K & K.km s$^{-1}$ & \\
\hline \hline
\\
H$^{13}$CN & E Region & 9.9 $\pm$ 3.7 & 11.4 & 40 $\pm$ 5 & 4.8 $\pm$ 0.5 & (8.2$^{+0.7}_{-1.0}$)E-03\\
 \\
 & NW Region & 11.73 $\pm$ 0.59 & 11.4 & 40 $\pm$ 5 & 8.2 $\pm$ 0.8 & (1.4$^{+0.1}_{-0.1}$)E-02\\
 \\
 & NE Region & 12.8 $\pm$ 1.2 & 11.4 & 40 $\pm$ 5 & 6.2 $\pm$ 0.6 & (1.0$^{+0.1}_{-0.8}$)E-02\\
\\
 & S Region  & 11.34 $\pm$ 0.39 & 11.4 & 40 $\pm$ 5 & 6.0 $\pm$ 0.6 & (1.00$^{+0.1}_{-0.1}$)E-02\\
 \\
 & FIR3 & 10.6 $\pm$ 2.7 & 11.4 & 40 $\pm$ 5 & 4.7 $\pm$ 0.5 & (6.3$^{+0.5}_{-0.7}$)E-03\\
 \\
 \hline
 \\
HC$^{15}$N & E Region & 9.6 $\pm$ 3.7 & 11.4 & 40 $\pm$ 5 & 0.36 $\pm$ 0.04 & (9.0$^{+0.5}_{-1.2}$)E-04\\
 \\
 & NW Region & 13.20 $\pm$ 0.38 & 11.4 & 40 $\pm$ 5 & 1.2 $\pm$ 0.1 & (2.4$^{+0.3}_{-0.3}$)E-03\\
 \\
 & NE Region & 12.5 $\pm$ 1.2 & 11.4 & 40 $\pm$ 5 & 1.2 $\pm$ 0.1 & (2.7$^{+0.3}_{-0.3}$)E-03\\
\\
 & S Region & 11.74 $\pm$ 0.89 & 11.4 & 40 $\pm$ 5 & 0.92 $\pm$ 0.09 & (2.0$^{+0.2}_{-0.2}$)E-03\\
 \\
 & FIR3 & 12.85 $\pm$ 1.5 & 11.4 & 40 $\pm$ 5 & 1.2 $\pm$ 0.1 & (2.4$^{+0.2}_{-0.2}$)E-03\\
 \\
\hline \hline
\\
$^{15}$NNH$^+$ & E Region & 0.80 & 11.67 $^{(a)}$ & 40.0$^{+5.0}_{-5.0}$ & 0.067 $\pm$ 0.007 & (1.2$^{+0.2}_{-0.2}$)E-03\\
\\
 & NW Region & 0.81$^{+0.02}_{-0.02}$ & 11.31$^{+0.07}_{-0.07}$ & 40.0$^{+0.0}_{-0.0}$ & 0.14 $\pm$ 0.01 & 0.1$ \pm$ 0.4\\
 \\
 & NE Region & 0.79$^{+0.02}_{-0.02}$ & 11.2$^{+0.2}_{-0.1}$ & 40.0$^{+0.4}_{-0.2}$ & 0.085 $\pm$ 0.009 & 0.1 $\pm$ 0.3\\
 \\
 & S Region  & 0.79$^{+0.02}_{-0.02}$ & 11.5$^{+0.2}_{-0.3}$ & 39.7$^{+0.4}_{-0.4}$ & 0.13 $\pm$ 0.01 & 0.1 $\pm$ 0.9\\
 \\
 \hline
 \\
N$^{15}$NH$^+$ & E Region & 0.80 & 10.30 $^{(a)}$ & 40.0$^{+5.0}_{-5.0}$ & 0.10 $\pm$ 0.01 & (2.1$^{+0.3}_{-0.3}$)E-03\\
\\
 & NW Region & 0.80$^{+0.02}_{-0.02}$ & 11.4$^{+0.2}_{-0.08}$ & 40.0$^{+0.2}_{-0.2}$ & 0.13 $\pm$ 0.01 & 1.4 $\pm$ 1.4\\
 \\
 & NE Region & 0.83$^{+0.04}_{-0.07}$ & 11.3$^{+0.1}_{-0.4}$ & 40.4$^{+0.6}_{-0.4}$ & 0.059 $\pm$ 0.006 & 0.1 $\pm$ 0.9\\
 \\
 & S Region & 0.80 & 10.67 $^{(a)}$ & 40.0$^{+5.0}_{-5.0}$ & 0.11 $\pm$ 0.01 & 0.1 $\pm$ 0.3\\
 \\
\hline
\\
N$_2$H$^+$ & E Region & 0.78$^{+0.04}_{-0.04}$ & 11.49$^{+0.02}_{-0.02}$ & 38.6$^{+1.0}_{-2.0}$ & 23 $\pm$ 2 & 3.20 $\pm$ 0.08\\
\\
 & NW Region & 0.75$^{+0.02}_{-0.02}$ & 11.11$^{+0.02}_{-0.02}$ & 38.4$^{+2.0}_{-2.0}$ & 36 $\pm$ 4 & 6.50 $\pm$ 0.06\\
 \\
 & NE Region & 0.97$^{+0.02}_{-0.02}$ & 11.18$^{+0.02}_{-0.02}$ & 43.8$^{+2.0}_{-6.0}$ & 30 $\pm$ 3 & 4.95 $\pm$ 0.06\\
 \\
 & S Region & 1.2$^{+0.00}_{-0.3}$ & 11.2$^{+0.3}_{-0.07}$ & 37.9$^{+7.0}_{-2.0}$ & 33 $\pm$ 3 & 5.550 $\pm$ 0.004\\
 \\
\hline
\end{tabular}
\\
\footnotesize{$^{(a)}$Due to low S/N, \rm{V}$_{LSR}$ is fixed to the value found from a single Gaussian fit of the strongest HF component using L-M fitting.}
\end{table*}
\section{Results}
\label{Results}
\begin{figure}[ht]
\centering
\captionsetup{justification=centering}
\includegraphics[width=230pt]{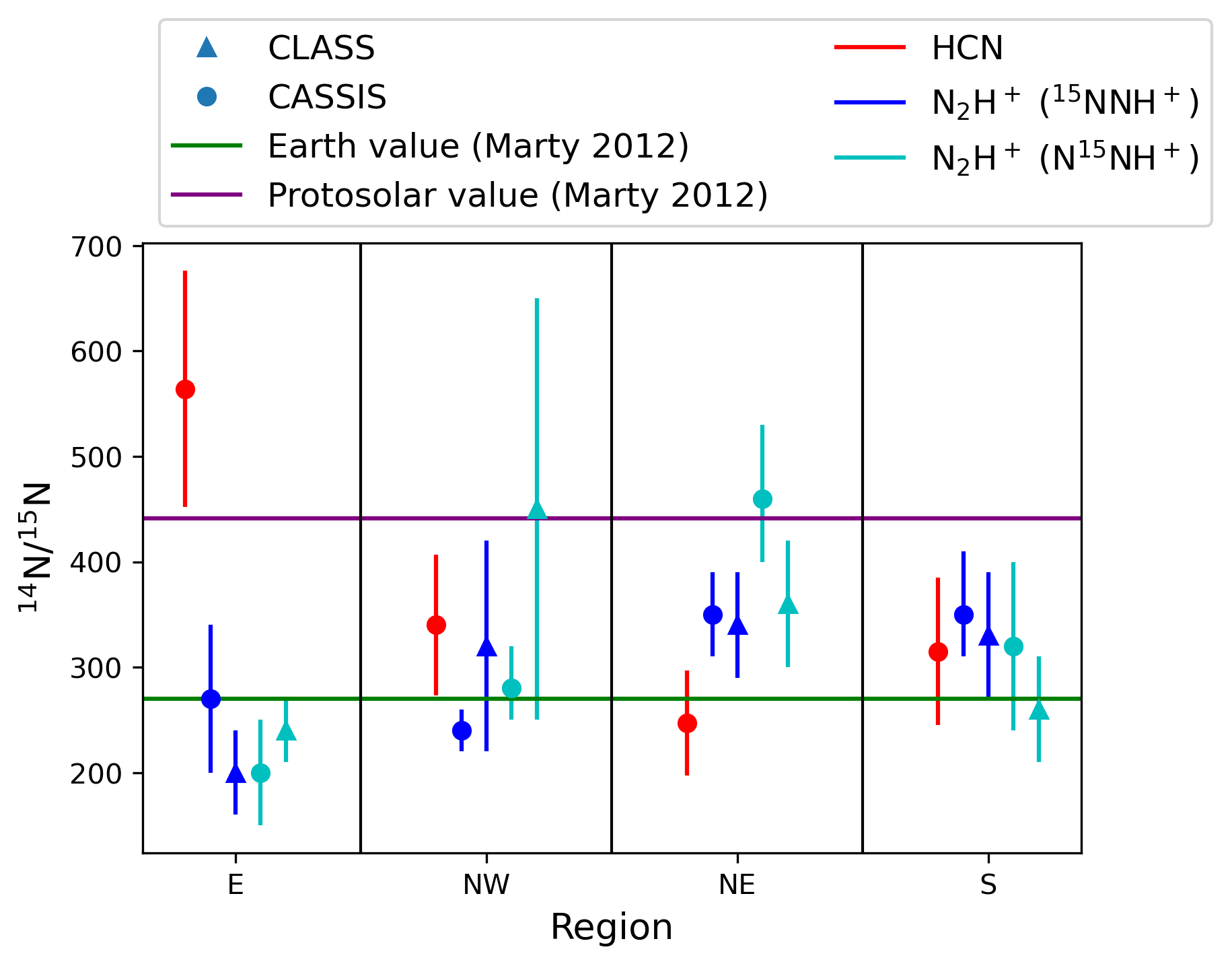}
\caption{Visual comparison of the $^{14}$N/$^{15}$N ratios obtained for HCN and N$_2$H$^+$ in the regions of OMC-2 FIR4, from both CASSIS and CLASS (in the latter case) and as calculated from the different $^{15}$N isotopologues (also in the latter case).}
\label{fig.ratios}
\end{figure}
\begin{figure}[ht]
\centering
\captionsetup{justification=centering}
\includegraphics[width=230pt]{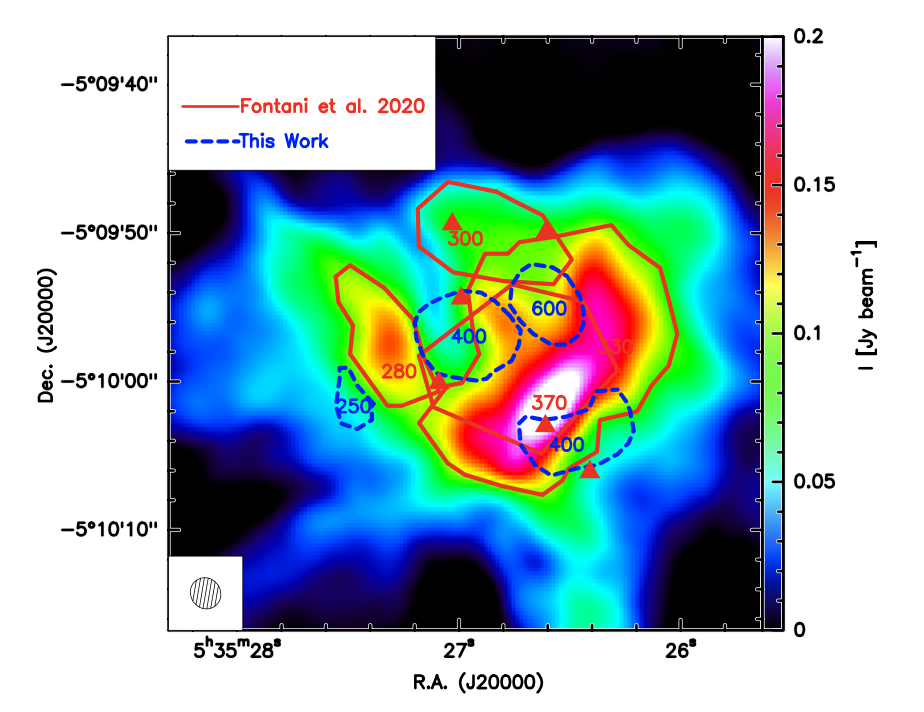}
\caption{Combined $^{14}$N/$^{15}$N ratios in N$_2$H$^+$ in regions defined by \citet{Fontani2020} (red) and in regions defined by our work (blue) overlaid onto N$_2$H$^+$ ($1-0$) emission from \citet{Fontani2020} (colour scale); red triangles show locations of continuum emission sources in FIR4 as identified by Neri et al. (in prep.). The wedge on the right indicates the range of flux density (Jy beam$^{-1}$). The ellipse in the bottom-left corner represents the ALMA synthesised beam.}
\label{fig_3}
\end{figure}
\noindent Figure \ref{fig.ratios} shows a visual comparison between all of the $^{14}$N/$^{15}$N ratios obtained during the course of our analysis, while Fig. \ref{fig_3} shows how the $^{14}$N/$^{15}$N ratios from our dataset and that described in \citet{Fontani2020} vary across the regions defined in each paper across OMC-2 FIR4. We now explore the results of our analysis of both the HCN and N$_2$H$^+$ datasets in detail.
\subsection{HCN}
\label{hcn}
As stated above, the column density values for H$^{13}$CN and HC$^{15}$N and the calculated $^{14}$N/$^{15}$N ratios can be found in Table \ref{Table_ncol_hcn}. It is clear from the values that there is an agreement in the $^{14}$N/$^{15}$N ratio in the NE, NW and S Regions but a higher value is obtained from the E Region. Meanwhile, FIR3 shows the lowest ratio. Overall, these results indicate that no enrichment or depletion in $^{15}$N is apparent in the inner sub-regions of FIR4 (i.e. the NW, NE and S Regions) analysed in this work. In fact, the derived $^{14}$N/$^{15}$N ratios are consistent with each other (within the uncertainties) as well as with the value obtained at low angular resolution by \citet{Kahane2018}: 270$\pm$50. The E Region shows the highest $^{14}$N/$^{15}$N ratio in HCN; however, it is not straightforward to justify this because, even though the 
region is the most external one and hence potentially most exposed to external UV photons, a variation due to selective photodissociation would produce a lower $^{14}$N/$^{15}$N ratio, which is at odds with our findings.
\begin{table*}[!h]
\captionsetup{justification=centering}
\caption{Values of column density obtained using CASSIS for H$^{13}$CN and HC$^{15}$N along with results of analysis using CASSIS for $^{14}$N/$^{15}$N in HCN. The values were calculated using average values of column densities taken at 35 K and 45 K (100 K and 150 K for FIR3).}
\label{Table_ncol_hcn}
\centering
\begin{tabular}{c c c c}
\hline  
Region & \multicolumn{2}{c}{\textit{N$_{TOT}$}} & $^{14}$N/$^{15}$N\\
 & \multicolumn{2}{c}{cm$^{-2}$} &\\
& H$^{13}$CN & HC$^{15}$N &\\
\hline \hline
\\
E Region & (1.9-2.3)$\times$10$^{13}$ & (1.7-2.1)$\times$10$^{12}$ & 452-676\\
\\
NW Region & (2.9-4.3)$\times$10$^{13}$ & (5.3-6.5)$\times$10$^{12}$ & 273-406\\
 \\
NE Region & (2.6-3.2)$\times$10$^{13}$ & (5.4-6.6)$\times$10$^{12}$ & 197-296\\
 \\
S Region  & (2.4-3.0)$\times$10$^{13}$ & (3.9-4.9)$\times$10$^{12}$ & 245-385\\
 \\
FIR3 & (1.9-2.4)$\times$10$^{13}$ & (5.0-6.3)$\times$10$^{12}$ & 151-240\\
\\
\hline
\hline
\end{tabular}
\end{table*}
\subsection{N$_2$H$^+$}\label{n2h+}
\begin{table*}[!h]
\captionsetup{justification=centering}
\caption{Column density values obtained using the two methods (CASSIS HFS and CLASS HFS) for $^{15}$NNH$^+$, N$^{15}$NH$^+$ and N$_2$H$^+$.}
\label{Table_ncol_n2h+}      
\centering          
\begin{tabular}{c c c c}
\hline  
Line & Region & \multicolumn{2}{c}{\textit{N$_{TOT}$}}\\
 & & \multicolumn{2}{c}{cm$^{-2}$}\\
 & & CASSIS HFS & CLASS HFS\\
\hline \hline
\\
$^{15}$NNH$^+$ & E Region & (3.0$^{+0.8}_{-0.7}$)$\times$10$^{11}$& (4.3$\pm$0.6)$\times$10$^{11}$\\
\\
 & NW Region & (5.0$^{+0.9}_{-0.9}$)$\times$10$^{11}$ & (5.3$\pm$0.7)$\times$10$^{11}$\\
 \\
 & NE Region & (3.1$^{+0.7}_{-0.7}$)$\times$10$^{11}$ & (3.8$\pm$0.5)$\times$10$^{11}$\\
 \\
 & S Region & (3.9$^{+1.0}_{-1.0}$)$\times$10$^{11}$ & (4.9$\pm$0.7)$\times$10$^{11}$\\
 \\
 \hline
 \\
N$^{15}$NH$^+$ & E Region & (4.0$^{+1.1}_{-1.0}$)$\times$10$^{11}$ & (3.7$\pm$0.3)$\times$10$^{11}$\\
\\
 & NW Region & (4.3$^{+1.0}_{-1.0}$)$\times$10$^{11}$ & (3.8$\pm$0.6)$\times$10$^{11}$\\
 \\
 & NE Region & (2.4$^{+0.7}_{-0.9}$)$\times$10$^{11}$ & (3.6$\pm$0.5)$\times$10$^{11}$\\
 \\
 & S Region & (4.25$^{+1.1}_{-1.1}$)$\times$10$^{11}$ & (6.1$\pm$0.8)$\times$10$^{11}$\\
 \\
\hline
\\
N$_2$H$^+$ & E Region & (8.2$^{+0.6}_{-0.6}$)$\times$10$^{13}$ & (8.9$\pm$0.9)$\times$10$^{13}$\\
\\
 & NW Region & (1.21$^{+0.09}_{-0.09}$)$\times$10$^{14}$ & (1.7$\pm$0.7)$\times$10$^{14}$\\
 \\
 & NE Region & (1.1$^{+0.1}_{-0.1}$)$\times$10$^{14}$ & (1.3$\pm$0.1)$\times$10$^{14}$\\
 \\
 & S Region & (1.4$^{+0.1}_{-0.2}$)$\times$10$^{14}$ & (1.6$\pm$0.2)$\times$10$^{14}$\\
 \\
\hline
\end{tabular}
\end{table*}
\begin{table*}[!h]
\captionsetup{justification=centering}
\caption{Results of CASSIS HFS and CLASS HFS analysis for $^{14}$N/$^{15}$N in N$_2$H$^+$. *In the E Region, a different second method was used instead of CLASS HFS due to the HFS structure being invisible (see Sect. \ref{n2h+}). The values were calculated using average values of column densities taken at 35 K and 45 K and can be seen in Table \ref{Table_ncol_n2h+}.}
\label{Table_2}      
\centering          
\begin{tabular}{c c c c c} 
\hline
 & \multicolumn{2}{c}{CASSIS HFS} & \multicolumn{2}{c}{CLASS HFS}\\
\hline  
Region & $^{14}$N/$^{15}$N ($^{15}$NNH$^+$) & $^{14}$N/$^{15}$N (N$^{15}$NH$^+$) & $^{14}$N/$^{15}$N ($^{15}$NNH$^+$) & $^{14}$N/$^{15}$N (N$^{15}$NH$^+$)\\
\hline \hline
\\
E Region* & 270$^{70}_{70}$ & 200$^{50}_{50}$ & 200$^{40}_{40}$ & 240$^{30}_{30}$ \\
\\
NW Region & 240$^{20}_{20}$ & 280$^{30}_{40}$ & 320$^{100}_{100}$ & 450$^{200}_{200}$\\
\\
NE Region & 350$^{40}_{40}$ & 460$^{70}_{60}$ & 340$^{50}_{50}$ & 360$^{60}_{60}$\\
\\
S Region & 350$^{60}_{40}$ & 320$^{80}_{80}$ & 330$^{60}_{60}$ & 260$^{50}_{50}$\\
\\
\hline
\end{tabular}
\end{table*}
The spectra of N$_2$H$^+$ and its $^{15}$N isotopologues have been fit according to the methods described in Sect. \ref{CLASS} and \ref{CASSIS}. The resulting values for $N_{\rm TOT}$ and $^{14}$N/$^{15}$N ratios are listed in Tables \ref{Table_ncol_n2h+} and \ref{Table_2}, respectively. The range of $^{14}$N/$^{15}$N that we obtain is 200-450. 

As one can see, in a similar fashion to the $^{14}$N/$^{15}$N ratios obtained from HCN, the ratios that we obtain from N$_2$H$^+$ show broad homogeneity across the three central regions. However, while our HCN results show the E Region to have a slightly higher $^{14}$N/$^{15}$N ratio, our results for N$_2$H$^+$ show the E Region to have a ratio that is still a little lower than most other regions but within error of the ratio found in at least one other region. There is therefore some agreement with previous research (in particular \citet{Fontani2020}, who analysed the same N$_2$H$^+$ data), which concluded that the $^{14}$N/$^{15}$N ratio showed little regional variation across the central regions of FIR4.

There are a few region-specific caveats in our analysis that should be highlighted. We now discuss these in more detail.

Regarding the S Region, as can be clearly seen in Fig. \ref{fig.spectra}, in most regions the F$_1=0-1$ component of N$_2$H$^+$ is not blended with the nearby HF components and shows a profile that can be well fitted with a single Gaussian, which also indicates that this particular HF component is likely optically thin, as assumed in Sect. \ref{CLASS} and \ref{CASSIS}. However, in the spectrum of the S Region this component shows two peaks. Interestingly, the N$^{15}$NH$^+$ line also shows two peaks in its main HF component. In both cases, this profile can be due to two partly overlapping velocity features. The $^{15}$NNH$^+$ line, instead, shows a single peak in the main component. To probe whether the profile observed in the F$_1=0-1$ component of N$_2$H$^+$ and in N$^{15}$NH$^+$ is due to multiple velocity features or self-absorption, we compared the velocities of the two peaks in the F$_1=0-1$ component of N$_2$H$^+$ ($1-0$) with those of N$^{15}$NH$^+$ and $^{15}$NNH$^+$. In both cases, the velocities of the peak(s) of the emission in the $^{15}$N isotopologues coincide with one (or both) of those seen in the isolated component of the N$_2$H$^+$ emission. Therefore, we can conclude that the double peak seen both in N$^{15}$NH$^+$ and in the isolated HF component of N$_2$H$^+$ is likely due to two velocity features along the line of sight.

In the NW Region, as stated before, the opacity derived in CLASS 
that resulted from using the HFS fitting routine is 1.4$\pm$1.4. With such a large error associated with a faint line, 
the opacity cannot be well constrained and this may be causing inaccuracies in our estimates performed assuming optically thin conditions. The slightly larger discrepancy between the ratios obtained using the two analysis methods (CASSIS and CLASS) in the NW Region (see Table \ref{Table_2}) can possibly be explained due to these potential inaccuracies.

In the case of the E Region, the line is so faint that only the main component of the HFS is visible in the $^{15}$NNH$^+$ and N$^{15}$NH$^+$ spectra. For this reason, the fit to the whole HFS cannot be performed. Therefore, we fitted the main HF component (the only one detected) with a single Gaussian with the resultant brightness temperature (\textit{T$_b$}) used to calculate the opacity ($\tau$) using Eq. \ref{eq.tau}. This opacity was in turn used to calculate the total column density (\textit{N$_{TOT}$}) in Eq. \ref{eq.lte}.
\section{Discussion}
\label{Discussion}
From our HCN results, we can see that there is a homogeneity in the $^{14}$N/$^{15}$N ratio across the central portion of FIR4, from which we extracted the NW, NE and S spectra, but not in the ratio of the E region. The difference in $^{14}$N/$^{15}$N ratio between the central three regions and the E region is possibly due to the E region being much further from the centre of FIR4, where conditions affecting $^{15}$N production may differ, including any possible outflow influence. A recent paper (\citet{Benedettini2021}) found that no nitrogen fractionation was occurring due to the shock in the bow shock of L1157; therefore we cannot and are not confirming that the different result in the E region is definitely due to an external outflow, especially given the lack of evidence. Rather, we are suggesting that this is a potential topic for future research using more sensitive observations.

The ratios obtained from N$_2$H$^+$ towards the central part of the source are also largely consistent among them and with those obtained from HCN (as explained previously, it is believed that small discrepancies seen in the results obtained from the two methods are more likely to be due to opacity effects than any true difference in the ratio; see Sect. \ref{n2h+}). However, in contrast to our HCN results for this region, which show the E Region to vary but not in the way that would be expected if photo-processes were important, our N$_2$H$^+$ ratio in the E region shows very little contrast with the ratios of the central regions, being the lowest of the four but still within uncertainty of at least one other region. 

Overall, the fact that in both HCN and N$_2$H$^+$ isotopologues we find no significant nitrogen fractionation effects confirms the previous findings of \citet{Fontani2020}, obtained only for N$_2$H$^+$ and analysed in different sub-regions: the nitrogen isotopic ratio in OMC-2 FIR4 in the two most abundant N-bearing species (after N$_2$) is not influenced significantly at core scales by a local change in physical conditions. In particular, in this work we have investigated for the first time the influence of an enhanced cosmic-ray ionisation rate (see Sect.~\ref{Results}) within the same source and found that this has no effect on the $^{14}$N/$^{15}$N ratio, a conclusion that confirms that of \citet{Fontani2020} and also of \citet{Benedettini2021} towards the protostellar bow-shock L1157-B1. The homogeneity that is present particularly across the central regions of FIR4 (NE, NW and S) in both HCN and N$_2$H$^+$ also goes towards the current theory that OMC-2 FIR4 is sufficiently embedded that selective photodissociation, which is expected to take place only on the external layers of dense cores exposed to external UV photons, is not present due to the high visual extinction of the source and hence does not affect the fractionation here.

\begin{figure}[ht]
\centering
\captionsetup{justification=centering}
\includegraphics[width=230pt]{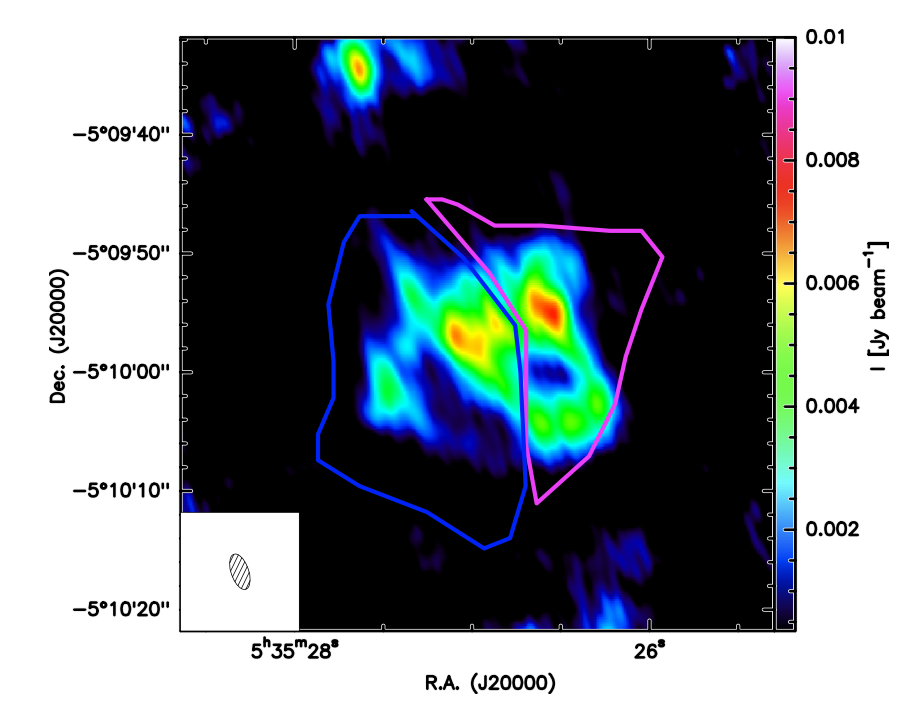}
\caption{Average emission map of H$^{13}$CN ($1-0$) overlaid with regions defined in \citet{Fontani2017} with suggested higher (blue) and lower (pink) ionisation rate regions. The wedge on the right indicates the range of flux density (Jy beam$^{-1}$). The ellipse in the bottom-left corner represents the NOEMA synthesised beam.}
\label{ionisation}
\end{figure}

The absence of significant variations in both species also indicates that selective photodissociation has no effect at the linear scales probed by our observations.
\citet{Colzi2019}, who studied the high-mass star-forming protocluster IRAS 05358+3543, measured a higher $^{14}$N/$^{15}$N ratio in N$_2$H$^+$ towards the diffuse region of the protocluster than towards the star-forming cores. Therefore, it was concluded that an enhanced $^{14}$N/$^{15}$N ratio in N$_2$H$^+$ occurs locally in the diffuse gas of star-forming regions. Despite the larger distance to the source compared to OMC-2 FIR4 (2~kpc vs. 393~pc) and thus the different linear scale, we can compare these results with our own by utilising the fact that the high ionisation rate extraction region (the blue polygon in Fig. \ref{ionisation}) overlaps with both the compact E and NE Regions as well as including the more diffuse regions. The consistency in the $^{14}$N/$^{15}$N ratio of N$_2$H$^+$ between our larger polygon and our more compact NE Region (both $\sim$400) 
suggests that, in this region, enrichment due to selective photodissociation caused by external UV photons is not occurring. 
This result could either be because there is an insufficient external UV field or because the emitting region is too embedded for external UV photons to penetrate and produce this effect. The same `non-variation' effect appears to be present also in the $^{14}$N/$^{15}$N ratios of HCN in OMC-2 FIR4 between the larger, more diffuse extraction region and the more compact NE Region (both $\sim$250).
\\
\\
\section{Conclusions}
\label{Conclusions}
In this work, within the framework of the Large Program SOLIS, we have mapped the H$^{13}$CN ($1-0$) and HC$^{15}$N ($1-0$) lines towards the protocluster OMC-2 FIR4 with the NOEMA interferometer. We have also analysed archival data obtained with the ALMA interferometer of the 1--0 lines of N$_2$H$^+$ and both its $^{15}$N isotopologues. For the first time, we compare at high angular resolution the $^{14}$N/$^{15}$N isotopic ratios in HCN and N$_2$H$^+$, two important N-bearing species that are expected to possibly behave differently in the enrichment or depletion of $^{15}$N. The results can be summarised as follows:
\begin{enumerate}
    \item H$^{13}$CN ($1-0$) and HC$^{15}$N ($1-0$) show spatial similarities in their emission, with H$^{13}$CN showing systematically stronger
    emission across all regions of the source. There are obvious peaks to both the H$^{13}$CN ($1-0$) and HC$^{15}$N ($1-0$) emission in the north-west, north-east and (to a lesser extent in HC$^{15}$N) south of the central area of emission, plus a peak farther to the east in an outer region that appears in H$^{13}$CN only. A noticeable absence of emission is also present in both species towards the centre of FIR4. There is also a peak in FIR3 in both H$^{13}$CN ($1-0$) and HC$^{15}$N ($1-0$) emission.
    \item The $^{14}$N/$^{15}$N ratio in HCN shows consistency across the different regions in which we have decomposed the emission, especially in the central part of FIR4 traced by regions NE, NW and S, while a higher isotopic ratio is found in the E region. This indicates that the main mechanism accepted as being behind enrichments of $^{15}$N in HCN (i.e. selective photodissociation) does not seem to play a role in these regions of the protocluster.
    \item The $^{14}$N/$^{15}$N ratios in N$_2$H$^+$ in the same regions as those selected for our HCN analysis are also obtained; these results show agreement with the HCN ratios in that there is consistency across the central regions while in the E region the $^{14}$N/$^{15}$N ratio from N$_2$H$^+$ slightly varies in the opposite direction compared to the ratio in HCN, although the errors are large. The ratios also show consistency across the central region of FIR4, in agreement with previous findings (\citet{Fontani2020}). 
    \item To further investigate our results, the $^{14}$N/$^{15}$N ratio in HCN in regions that have previously shown a different 
    cosmic-ray ionisation rate are analysed and found to be consistent with each other, further cementing our conclusion that the $^{14}$N/$^{15}$N ratio is not influenced by local changes in the physical properties across the centre of FIR4.
    \item Measuring the $^{14}$N/$^{15}$N ratio of N$_2$H$^+$ in a region encompassing the diffuse envelope and the E and NE Regions, as well as regions external to both suggests consistency between the diffuse envelope and the NE Region, seemingly confirming a `non-variation' effect and once again indicating that selective photodissociation seems to be irrelevant in these regions.
\end{enumerate}
It is clear from our results and conclusions that our picture of the evolution of the $^{14}$N/$^{15}$N ratio in protostellar environments, as well as on the origin of $^{15}$N enrichments in SS small bodies, is still incomplete. More collaboration between theory and observations is needed here in order to resolve these questions. More specifically, in the case of FIR4, the observation of higher transitions will allow us to better constrain parameters such as $T_{\rm ex}$ and $N_{\rm TOT}$ in order to create more accurate models to compare with our observations in both LTE and non-LTE. 
\section*{Acknowledgements}
This project has received funding from (i) the European Union Horizon 2020 research and innovation programme under the Marie Skłodowska-Curie grant agreement No 811312 for the Project "Astro-Chemical Origins, (ACO) and (ii) the European Research Council (ERC) under the European Union Horizon 2020 research and innovation programme, for the Project "The Dawn of Organic Chemistry" (DOC), grant agreement No 741002. This paper makes use of the following ALMA data: ADS/JAO.ALMA\#2016.0.00681.S. ALMA is a partnership of ESO (representing its member states), NSF (USA) and NINS (Japan), together with NRC (Canada), MOST and ASIAA (Taiwan) and KASI (Republic of Korea), in cooperation with the Republic of Chile. The Joint ALMA Observatory is operated by ESO, AUI/NRAO and
NAOJ. We thank Laura Colzi for useful discussions and are also grateful to the IRAM staff for help with data calibration.
\bibliographystyle{aa}
\bibliography{42147corr}
\newpage
\onecolumn
\appendix
\section{Spectra}\label{Spectra}
\begin{figure}[h]
\centering
\captionsetup{justification=centering}
\includegraphics[width=500pt]{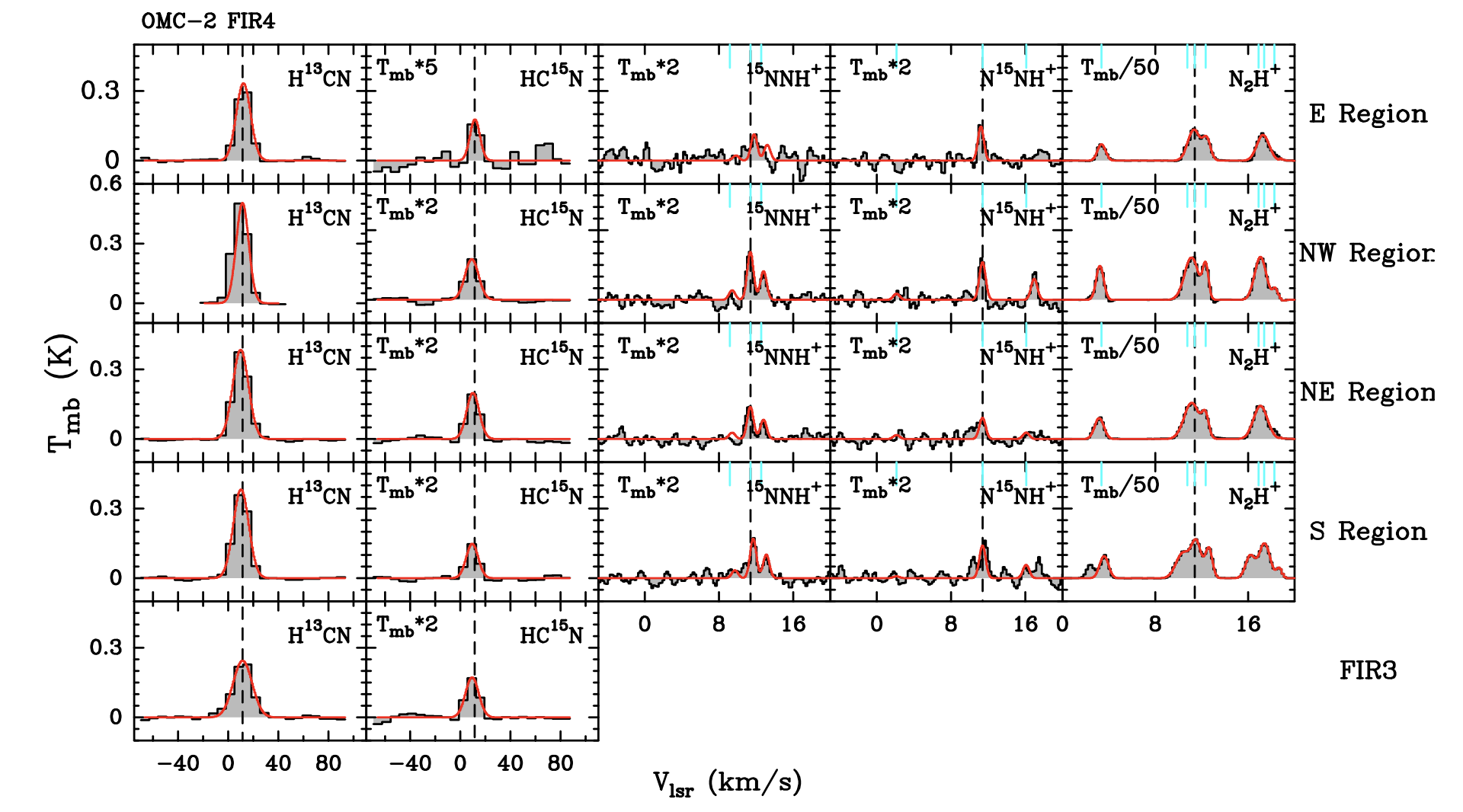}
\caption{Spectra extracted for our study in (left to right) H$^{13}$CN ($1-0$), HC$^{15}$N ($1-0$), $^{15}$NNH$^+$ ($1-0$), N$^{15}$NH$^+$ ($1-0$) and N$_2$H$^+$ ($1-0$); from top to bottom, the spectra extracted in the E, NW, NE, S and FIR3 regions. The CASSIS fits are superimposed in red. The dashed black line represents the \rm{V}$_{LSR}$ value for this source, which is 11.4 km s$^{-1}$. The blue lines represent the location of the HFS components for N$_2$H$^+$ and its $^{15}$N isotopologues.}
\label{fig.spectra}
\end{figure}
\end{document}